\documentclass[prb,twocolumn,preprintnumbers,amsmath,amssymb]{revtex4-1}

\usepackage{graphicx}
\usepackage{dcolumn}
\usepackage{bm}
\usepackage{booktabs,threeparttable}
\usepackage{subfigure}
\usepackage{textcomp}

\makeatletter 
\def\@hangfrom@section#1#2#3{\@hangfrom{#1#2#3}}
\makeatother

\makeatletter        
\def\@biblabel#1{[#1]}
\makeatother

\makeatletter
\newcommand\figcaption{\def\@captype{figure}\caption}
\newcommand\tabcaption{\def\@captype{table}\caption}
\makeatother

\usepackage{color}

\begin{document}
\title{Chemical Ordering and Crystal Nucleation at the Liquid Surface: A Comparison of $\rm{Cu}_{50}\rm{Zr}_{50}$ and $\rm{Ni}_{50}\rm{Al}_{50}$ Alloys}
\author{Chunguang Tang$^{1,2}$ and Peter Harrowell$^{2}$}
\affiliation{$^1$School of Materials Science and Engineering, University of New South Wales, NSW 2052, Australia; $^2$School of Chemistry, The University of Sydney, NSW 2006, Australia}

\begin{abstract}
We study the influence of the liquid-vapor surface on the crystallization kinetics of supercooled metal alloys. While a good glass former, $\rm{Cu}_{50}\rm{Zr}_{50}$, shows no evidence of surface enhancement of crystallization, $\rm{Ni}_{50}\rm{Al}_{50}$ exhibits an increased rate of crystallization due to heterogeneous nucleation at the free liquid surface. The difference in the compositional fluctuations at the interface is proposed as the explanation of the distinction between the two alloys. Specifically, we observe compositional ordering at the surface of $\rm{Ni}_{50}\rm{Al}_{50}$ while the $\rm{Cu}_{50}\rm{Zr}_{50}$ alloy only exhibits a diffuse adsorption of the Cu at the interface. We argue that the general difference in composition susceptibilities at planar surfaces represents an important factor in understanding the difference in the glass forming ability of the two alloys.

\end{abstract}

\maketitle

\section{Introduction} The glass forming ability (GFA) of a molten alloy is established, in practice, by the stability of the melt with respect to heterogeneous nucleation. Walls, foreign particles and the liquid-vapour interface typically provide the kinetically dominant pathway to crystallization. While the first two sources of heterogeneous nucleation can be avoided by careful design, the liquid-vapour interface is generally unavoidable, an intrinsic potential source of crystallization nucleation sites and, consequently, a useful probe as to the  chemical origin of glass forming ability in alloys. In this paper we shall report on molecular dynamics (MD) simulations of the influence of the liquid surface on the stability of two metal alloys: one a good glass former ($\rm{Cu}_{50}\rm{Zr}_{50}$) and the other, a poor glass former, ($\rm{Ni}_{50}\rm{Al}_{50}$).

Previously~\cite{nature-mat}, we have established that the model alloy $\rm{Ni}_{50}\rm{Al}_{50}$ exhibits significantly faster crystallization rates in the bulk liquid and faster crystal growth rates than does the model $\rm{Cu}_{50}\rm{Zr}_{50}$, in agreement with the experimental observation. Understanding the difference in GFA between these two alloys is important because the $\rm{CuZr}$ alloys exhibit an anomalously high GFA for a binary alloy, making them an ideal system to address the basic causes of slow crystallization. While providing a representative example of a poor glass forming alloy, $\rm{Ni}_{50}\rm{Al}_{50}$, shares a number of similarities with  $\rm{Cu}_{50}\rm{Zr}_{50}$, including composition, similar atomic radius ratios, the same primary crystalline phase and liquid diffusion constants of similar magnitude. The explanation of their different values of GFA, therefore, represents a non-trivial puzzle, one central to understanding the factors that control crystallization kinetics.

There is a considerable literature on the role of the liquid-vapour surface as a site of heterogeneous crystal nucleation. Surface crystallization of supercooled water has been identified as the principal route for the formation of ice clouds~\cite{reiss}. Studies on the surface crystallization of molecular glass formers have demonstrated that the crystals can grow outward from the liquid~\cite{ediger}. In the case of surface crystallization of liquid metals~\cite{koster}, much of the research has concentrated on the important influence of surface oxidation in the chemical generation of nucleation centres. In the absence of oxygen, surfaces offer a number of physical advantages for nucleation: i) a reduction in the surface free energy penalty~\cite{surface-fe}, ii) reduction of stress associated ~\cite{stress} and iii) composition fluctuations due to surface adsorption~\cite{gibbs}. If an alloy is to be a good glass former, all of these possible avenues of enhanced nucleation at the liquid surface must be suppressed.

In the following we shall examine the kinetics of crystallization in simulations of thick films of supercooled liquids of NiAl and CuZr bounded by free surfaces and the associated compositional structure at these surfaces.

\section{Model and Algorithm}

The simulation of the liquid surface has been carried out as follows. First, a bulk cubic simulation cell containing 5488 atoms (2744 atoms of each species) was heated from 300 K up to 2300 K and then cooled to 1600 K (for NiAl) or 1400 K (for CuZr) at the speed of 100 K per 200 ps. The melting points of these two alloys were previously estimated by simulation~\cite{nature-mat} to be 1530 K and 1340 K, respectively. The free surfaces were then created by extending the simulation cell boundaries along one Cartesian direction for equilibrated liquid atomic configuration at 1600 K (for NiAl) and 1400 K (for CuZr) while leaving the particle positions unaltered to effectively create a vacuum layer about 30 $ $\r{A}$ $ thick. With periodic boundaries this procedure results in cleaving the liquid to create two surfaces. Time is measured from the creation of the surfaces. The
NiAl liquid plus surface was then relaxed at 1600 K for 3 ns before quenched and relaxed at 900 K, a temperature close to the `nose' of its time-temperature-transformation curve. For NiAl, we also relaxed the bulk system at 900 K, after equilibrated at 1600 K for about 4 ns, for comparison with the liquid plus surface. Using the measured enthalpy of fusion, we estimate that a temperature of 700 K will produce a thermodynamic driving force for the CuZr crystallization equivalent to that experienced by the NiAl alloy at 900 K. The CuZr liquid plus surfaces was quenched to a range of temperatures between 650 K and 1100 K after being equilibrated at 1400 K for 10 ns.  All the bulk simulations in this work were carried out at constant pressure (zero pressure) and temperature while all the simulations of the surface were carried out at constant volume (i.e liquid + vapor)  and temperature. In addition to the above simulations, we also computed the surface potential energies of pure elements in face centered cubic and amorphous phases by comparing the energies of a slab and its bulk phase at zero temperature. For the surface energy of amorphous phase, a series of configurations were used to obtain the average value. LAMMPS~\cite{LAMMPS} MD simulation software was used. The atomic interactions were modelled by the embedded atom method (EAM) potential proposed by Mishin et al.~\cite{Mishin2002} for NiAl and by Mendelev et al.~\cite{Mendelev2009} for CuZr, respectively.

\begin{figure}[!t]
\centering
\includegraphics[width=3in]{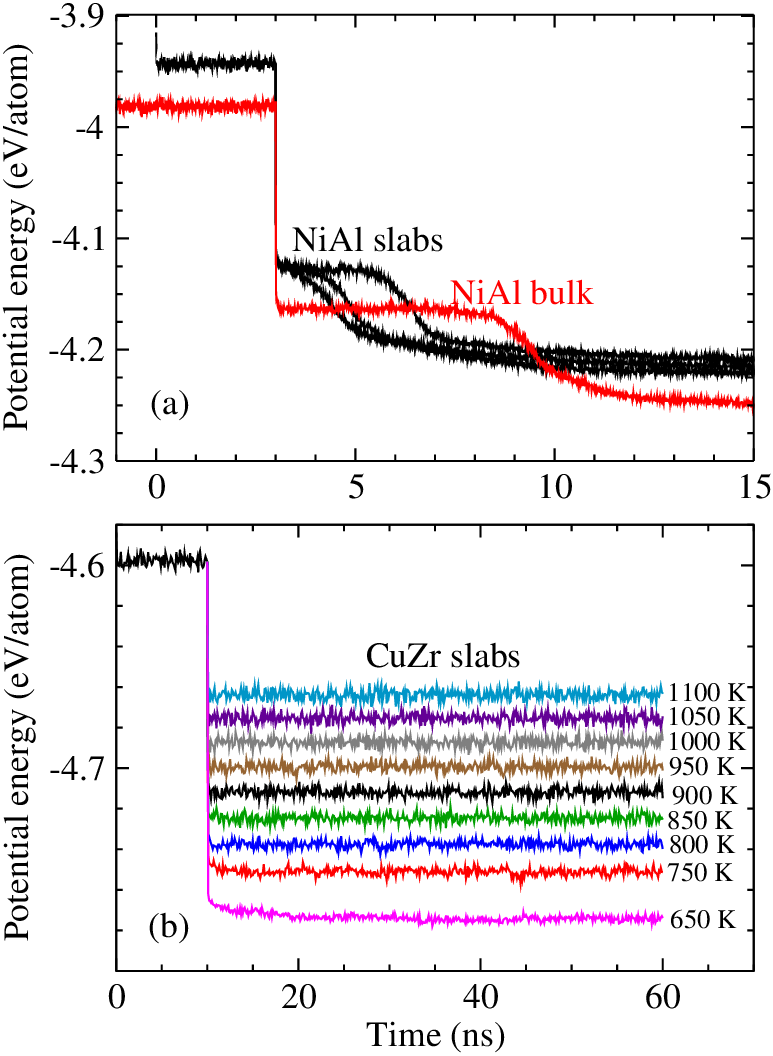}
\caption{(a) Potential energy of NiAl quenched from 1600 K to 900 K. Curves in black are for surface model (three independent runs) and curve in red is for bulk model. (b) Potential energy of CuZr slabs quenched from 1400 K to various temperatures. For 1050, 850, and 650 K, the simulations have been extended to 100 ns (not shown) and no crystallization was observed.}
\label{fig:pe}
\end{figure}

\section{Crystallization at the Liquid-Vapour Surface}


Crystallization can be conveniently monitored by recording the potential energy as a function to time. A decrease in the potential energy following an incubation interval over which the energy was stationary is characteristic of crystal nucleation. As presented in Fig. \ref{fig:pe}, we find that the $\rm{Ni}_{50}\rm{Al}_{50}$ alloy exhibits rapid crystallization in the presence of the liquid surface, significantly faster than that observed in the bulk liquid under the same conditions. For the bulk liquid, the incubation time is about 5 ns, and for the liquid with surface the observed longest incubation time is about 2.5 ns.  In contrast, we observe no sign of crystallization in the $\rm{Cu}_{50}\rm{Zr}_{50}$ alloy in the presence of the liquid surface at any of the temperatures studied.

\begin{figure}[!t]
\centering
\includegraphics[width=3.55in]{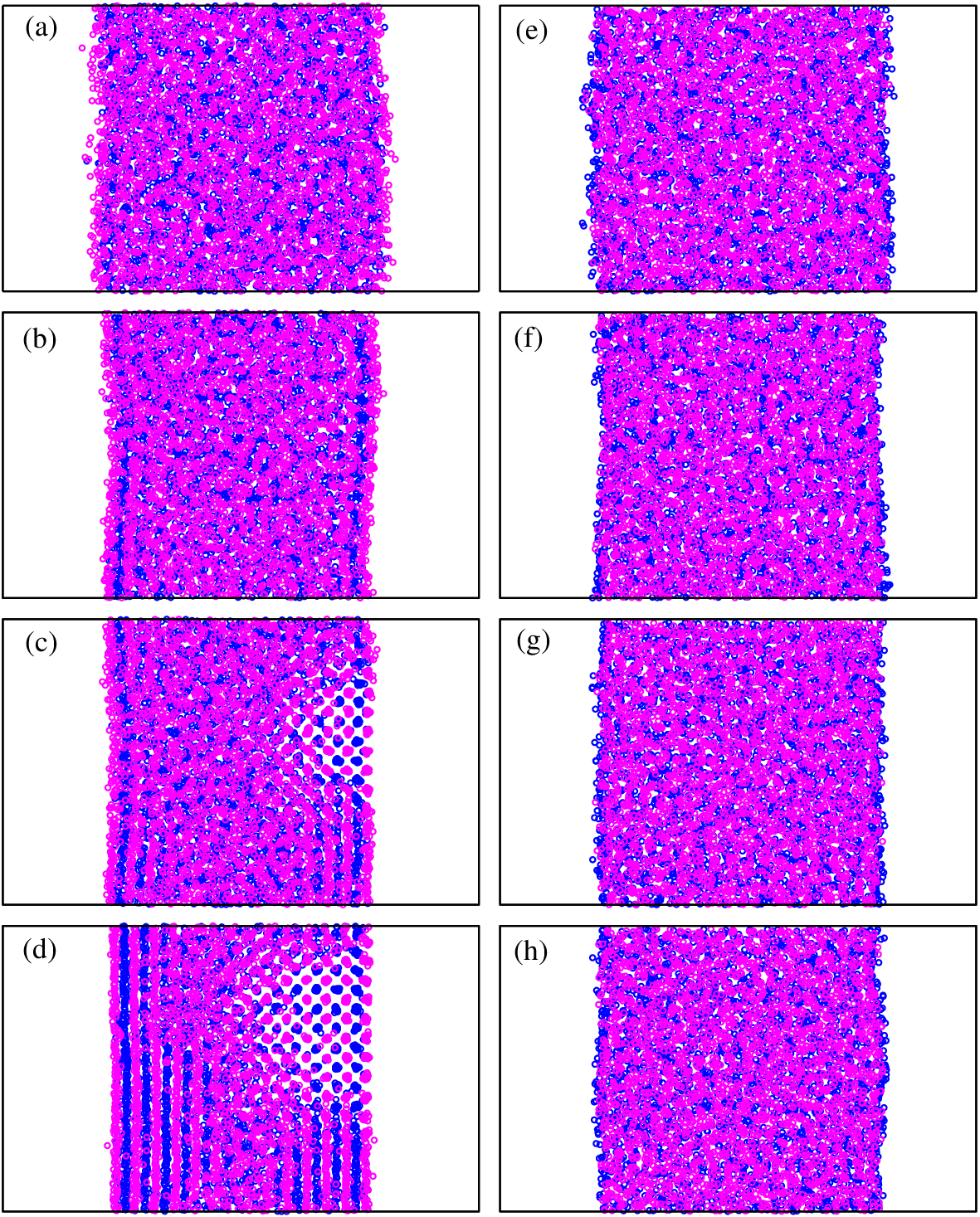}
\caption{A series of snapshots of the crystallization process of a NiAl slab (a-d) and a CuZr slab (e-h), with the positions of atoms projected on a plane through the slab (with normal parallel to the surface). For NiAl, (a) t=3 ns and T=1600 K; (b) 3.5 ns, 900 K; (c) 4 ns, 900 K; (d) 5 ns, 900 K. For CuZr, (e) 10 ns, 1400 K; (f) 20 ns, 700 K; (g) 40 ns, 700 K; (h) 60 ns, 700 K. Ni/Cu in blue and Al/Zr in magenta. }
\label{fig:snapshots}
\end{figure}

\begin{figure*}[!t]
\centering
\includegraphics[width=6in]{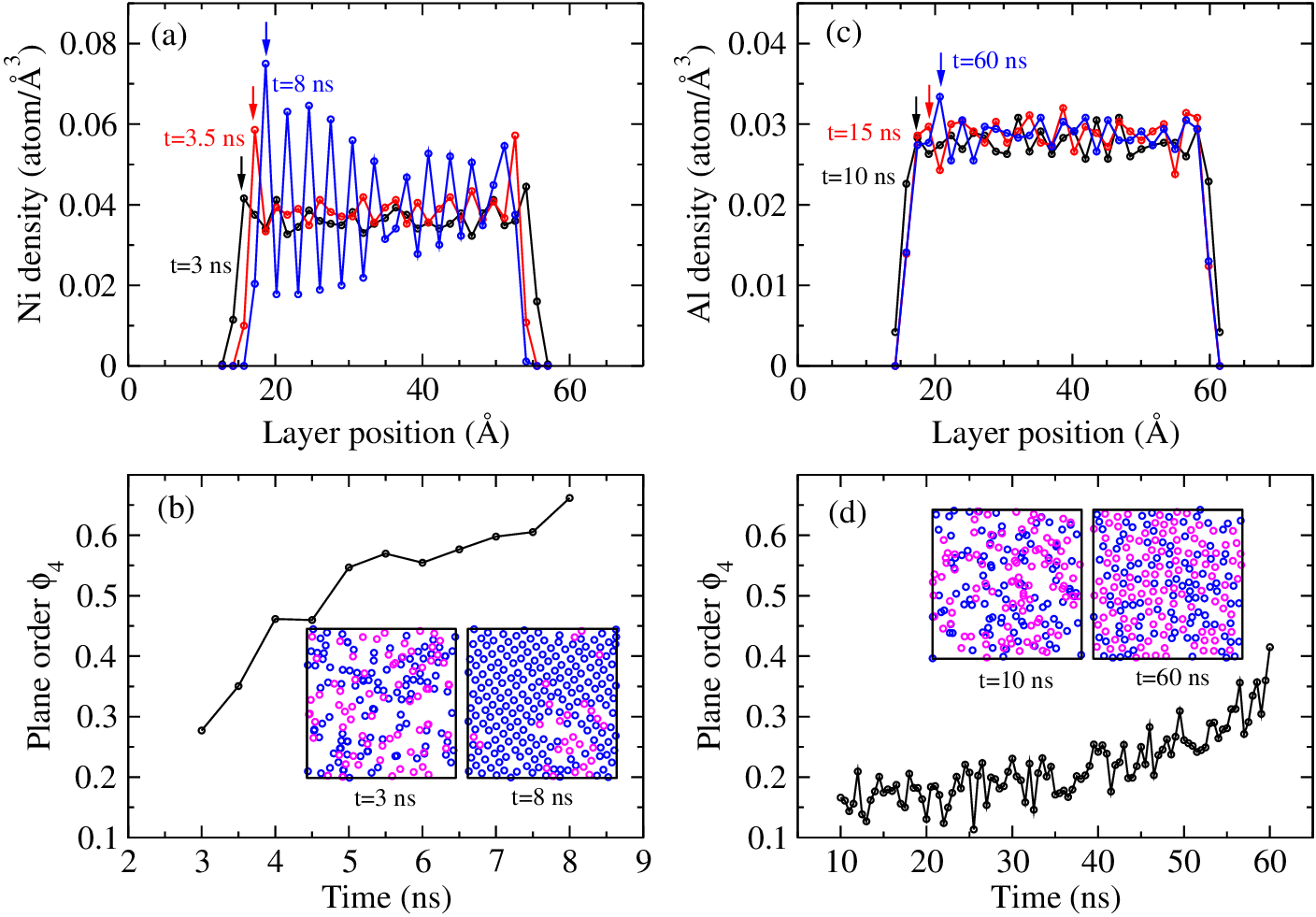}
\caption{(a) Ni density as a function of layer position along surface normal and time at 900 K. The layer thickness is 1.475 $ $\AA$ $. (b) In-plane order parameter $\phi_4$ of the sub-surface layer (see text), as is identified by the arrows in (a). The cutoff of bond length is 2.95 $ $\AA$ $, roughly the B2 lattice parameter at 900 K. The insets show the sub-surface layer structure projected on plane parallel to the surface. (c) Zr density at 700 K, with layer thickness of 1.63 $ $\AA$ $. (d) Order parameter $\phi_4$ corresponding to (c), with bond cutoff being 3.26 $ $\AA$ $. }
\label{fig:plane-order}
\end{figure*}

A sequence of snapshots of the evolution of order in the ${\rm{Ni}_{50}\rm{Al}_{50}}$ slab reveal that crystallization is nucleated at the surface (see Fig. \ref{fig:snapshots}). The critical nucleus appears to take the form of the hemisphere of crystal, just as assumed in the classical theory of heterogeneous nucleation~\cite{surface-fe}. These configurational ``snapshots'' also provide a key clue as to the mechanism of this heterogeneous nucleation: the surface of the NiAl alloys shows clear compositional layering, similar to the layering of the (100) plane of the B2 crystal phase, once the liquid film has been cooled to 900K. It is within this compositional bilayer that the surface-mediated nucleation takes place. As is shown in Fig.~\ref{fig:snapshots}, no such layering is evident at the surface of the CuZr alloy, at least for the 60ns we can run the simulation.

\begin{figure} [!t]
\centering
\includegraphics[width=3.in]{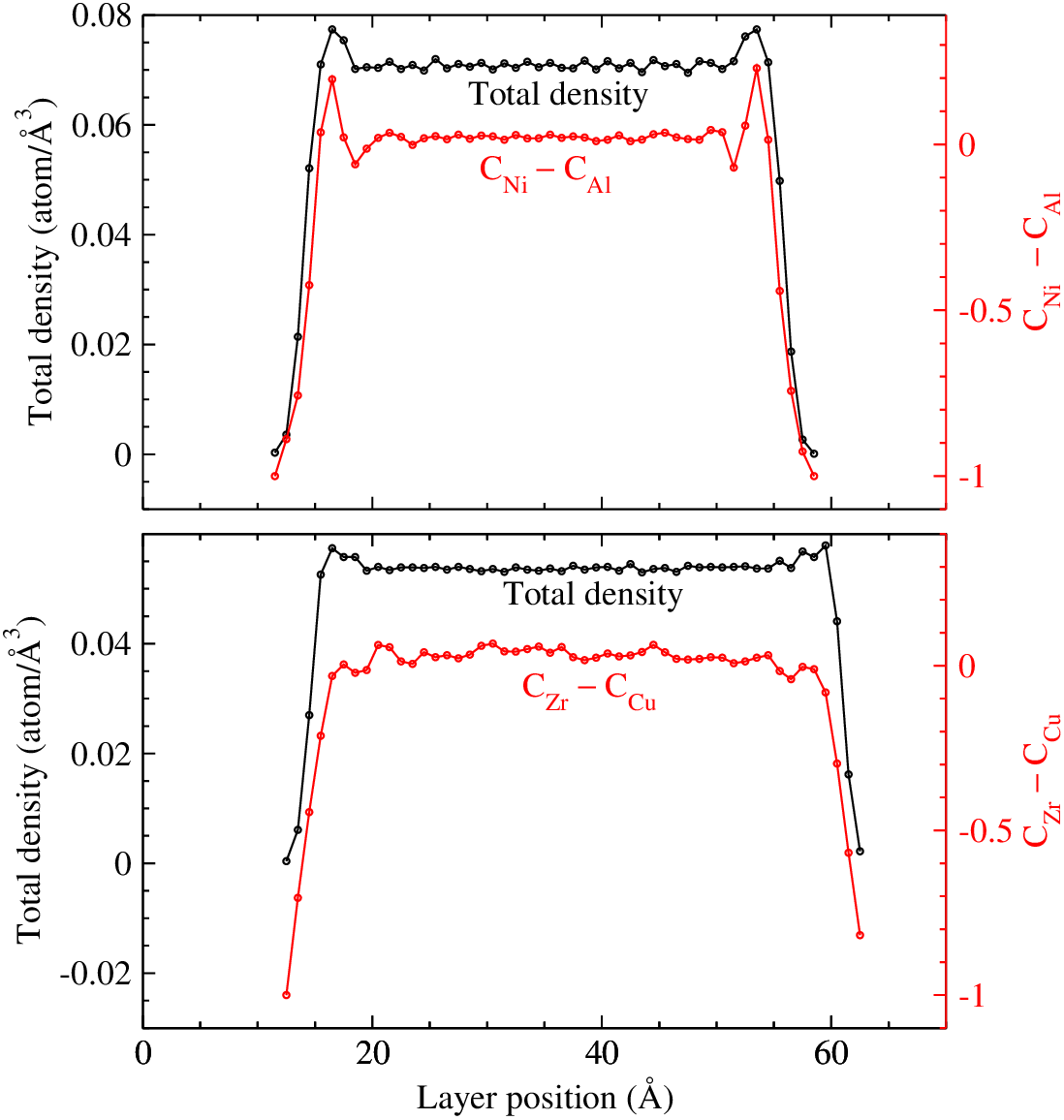}
\caption{(a) The total density profile and the profile of the composition difference $C_{Ni}-C_{Al}$ through the liquid slab at T = 1600 K. (b) The total density profile and the profile of the composition difference $C_{Zr}-C_{Cu}$ liquid slab at T = 1400 K. Note the preferential surface adsorption of Al and Cu, respectively, and the peak in the Ni density, indicative of Ni layering just under the liquid surface.}
\label{fig:density}
\end{figure}

Previous studies~\cite{greer2006,greer2010} have highlighted the point that layering of a liquid at a surface does not automatically imply enhanced crystal nucleation kinetics;  the degree to which the surface can promote in-plane order is also important. In the case of the B2 crystal structure, the in-plane order is that of a square lattice with an order parameter $\phi_4$, which for atom $i$ defined here as $\overline{\rm{cos}^2(4\theta)}$, where $\theta$ is the bond angle formed by atom $i$ and its two nearest neighbour atoms and the overline indicates the average over all bond angles of atom $i$. For a layer of ideal B2 crystal, one obtains $\phi_4$=1. The evolution of both layering and in-plane order is presented in Fig. \ref{fig:plane-order}. The atoms used for $\phi_4$ calculation are confined within the sub-surface layer in each alloy. We see that the growth of periodic density fluctuations of Ni atoms along the surface normal in the NiAl mixture is accompanied by a steady increase in the in-plane order as measured by $\phi_4$. In the case of CuZr, we find that there is a slow but systematic increase in both the compositional layering and the in-plane order without actually achieving any ordered crystal within the 60ns limit on the run time.

\begin{table}
\caption{Surface energies of pure elements in face centered cubic and amorphous phases (see text). The surface energies of amorphous phases are averaged over 40 amorphous configurations. Experimental surface energies\cite{Tyson1977} for Ni, Al, Cu, and Zr (in hexagonal close-packed phase) are 0.15, 0.07, 0.11, and 0.12 eV/$ $\AA$ $$^2$, respectively.}
\tabcolsep=12pt
\begin{tabular} {ccc}
  \hline
  \hline
Element$_{\rm{phase}}$  &  face & surface energy (eV/$ $\AA$ $$^2$)  \\
\hline
Ni$_{\rm{fcc}}$ & (100) & 0.08  \\
Al$_{\rm{fcc}}$ & (100) & 0.06   \\
Ni$_{\rm{fcc}}$ & (110) & 0.12 \\
Al$_{\rm{fcc}}$ & (110) & 0.07  \\
Ni$_{\rm{amor}}$ & ~ & 0.11$\pm$0.003    \\
Al$_{\rm{amor}}$ & ~ & 0.06$\pm$0.001   \\
\hline
Cu$_{\rm{fcc}}$ & (100) & 0.07  \\
Zr$_{\rm{fcc}}$ & (100) & 0.09   \\
Cu$_{\rm{fcc}}$ & (110) & 0.07  \\
Zr$_{\rm{fcc}}$ & (110) & 0.09   \\
Cu$_{\rm{amor}}$ & ~ & 0.06$\pm$0.004    \\
Zr$_{\rm{amor}}$ & ~ & 0.09$\pm$0.003  \\
\hline
\hline
\end{tabular}
\label{tab:surfe}
\end{table}
\vspace{20pt}

\section{Structure at the Surface of the Liquid Alloys}
In Fig.~\ref{fig:density} we have plotted the average profile of the total density and the composition differences, $C_{Ni}-C_{Al}$ and $C_{Zr}-C_{Cu}$ respectively, for the two alloys. We find that both surfaces show clear evidence of species segregation (i.e. large deviations of the composition difference from zero). Al and Cu are selectively adsorbed to the surfaces in the two alloys, consistent with their lower surface energies as indicated in Table~\ref{tab:surfe}. Here we have defined a surface energy as the energy increase resulting from cleaving a bulk material at 0K followed by an energy minimization, i.e. $(E_{\rm{c}}-E_{\rm{b}})/(2A)$, where $E_{\rm{c}}$ and $E_{\rm{b}}$ are the energies of the relaxed cleaved and bulk systems, respectively, and 2$A$ is the area of the two created surfaces. In the case of NiAl, we also observe a substantial second segregation peak of Ni that appears adjacent to the surface layer. This second layer is so large that it is accompanied by a clear peak in the total density. No such secondary segregation is observed in the CuZr liquid although there is a weak peak in the total density.

To recap, the two liquid alloys, NiAl and CuZr, exhibit dramatically different crystallization kinetics while exhibiting only small differences in structure~\cite{small}. The term 'small' is used here in the sense that the we have not been able to establish any compelling link between the observed differences in structure and crystallization kinetics. In this paper, we have demonstrated a clear difference between the two liquids (i.e. the presence or absence of compositional layering at the liquid surface) and shown that this difference directly influences crystallization kinetics (i.e. through the capacity of the surface to act as a heterogeneously nucleate crystals). It follows, therefore, that the presence or absence of compositional ordering at the liquid surface represents a useful intrinsic signature of the glass forming ability of an alloy. There is x-ray scattering data on layering at the liquid interface for pure metals~\cite{metals} and metals mixed with silicon or germanium~\cite{silicon1,silicon2}. While the experimental data for non-metallic liquids is scarcer there are reports of layering at the surface of liquid silicon and germanium~\cite{germ}. A number of these studies~\cite{silicon1,sutter} include evidence that crystallization is nucleated at the free surface.

Beyond an empirical signature of glass forming ability, it seems fruitful to ask whether the physical factors responsible for layering at a liquid surface are also those responsible for establishing the difference in glass forming ability in these alloys. In making this proposal, we are fully aware that homogeneous crystal nucleation has no explicit dependence on the nature of the liquid surface. Our reasoning in raising this proposition is as follows. i) Previously~\cite{nature-mat}, we established that the large difference in crystal growth rates between NiAl and CuZr was associated with the difference in chemical ordering at the liquid-crystal interface. The NiAl alloy exhibits an extended compositional layering at the interface while the CuZr liquid does not. ii) In the absence of striking differences in the average structural correlations of the two liquids, it is plausible to suggest that the important difference lies in their response to a perturbation. The liquid-vapor interface can be viewed as such a perturbation, one whose connection with crystallization is that both represent extended density variations. Exactly what kind of perturbation does the liquid metal surface represent? Rice and coworkers~\cite{rice} presented an explanation of a non-monotonic density profile through the interface based on the many body effects associated with a metal-nonmetal transition through the liquid-vapor interface. Celestini et al~\cite{celestini} subsequently found layers at the surface of liquid gold simulated using a many-body potential similar to the EAM potentials used in this paper. It is not all clear that potentials of the Embedded Atom type properly describe the metal-nonmetal invoked by Rice and yet the model generates density oscillation in the interface. The resolution of this puzzle may have been provided by Chac\'on et al~\cite{tarazona-2001}, who argued that {\it all} simple liquids should exhibit surface layering as long as the melting point is below $\approx 0.2T_{c}$, where $T_{c}$ is the gas-liquid critical temperature. These authors found that while a Lennard-Jones liquid exhibited a monotonic interface, the analogous liquid with a softer core (and, hence, lower freezing point) exhibited a layered surface. According to ref.~\cite{tarazona-2001}, oscillations occur in some metal surfaces, not because of a metal-nonmetal transition, but simply because a number of metals do have sufficiently low melting points to stabilize this partial order. In Fig. \ref{fig:density}, we find that the NiAl surface has significantly larger peaks in the density and the concentration difference than does CuZr. Is it that the surface stabilizes a density layer that, in turn drives the composition fluctuation, or does strong surface segregation at the surface drive the density build up? In the absence of any compelling evidence that the temperatures that we have used are sufficiently 'low' (as defined in ref.~\cite{tarazona-2001}) we have suggested here that the latter scenario, i.e. that in which composition fluctuations drive the surface ordering, is responsible for the difference in surface ordering between the two liquids and, by extension, is related to their overall difference in glass forming ability.

\section{Conclusion}

In conclusion, we have demonstrated that the difference in the glass forming ability already established for the simulated alloys $\rm{Ni}_{50}\rm{Al}_{50}$ and $\rm{Cu}_{50}\rm{Zr}_{50}$  in the bulk liquid is retained when considering the influence of heterogeneous nucleation of the crystal at the liquid surface. We demonstrate that the NiAl alloy undergoes heterogeneous nucleation as a result of the short wavelength chemical order induced at the surface. The CuZr surface exhibits no such chemical layering, despite the fact that Cu is selectively adsorbed at the surface. Kaban et al ~\cite{kaban} have argued that the difference in chemical ordering (in their case between $\rm{Ni}_{64}\rm{Zr}_{36}$ and $\rm{Cu}_{65}\rm{Zr}_{35}$) can play a central role in determining the glass forming ability of an alloy. It is reasonable to suggest that this tendency for chemical order may, if sufficiently strong, result in the oscillatory partial densities in the liquid-vapour interface that we have observed here in the simulated  $\rm{Ni}_{50}\rm{Al}_{50}$ alloy and, hence, to enhanced heterogeneous crystal nucleation at the liquid surface. What is intriguing about the results reported here is that this tendency of the NiAl alloy to form chemical layers at the liquid surface is very similar to the analogous ordering previously observed at the crystal-liquid interface,\cite{nature-mat} a feature that was associated with the significant difference in the crystal growth rates in the NiAl and CuZr alloys. The recurrence of this susceptibility to chemical order as a key determinant of the crystallization kinetics of the alloy underscores the key role composition fluctuations play in determining the glass forming ability of metal alloys.

{\bf{Acknowledgements}}
We acknowledge financial support from the Australian Research Council. CT would particularly like to thank the Australian Research Council for the DECRA Fellowship (grant no. DE150100738). Part of this research was undertaken on the NCI National Facility in Canberra, Australia, which is supported by the Australian Commonwealth Government.

\end{document}